\documentclass[runningheads,a4paper]{llncs}

\usepackage{amssymb}
\usepackage{graphicx}

\usepackage{url}
\urldef{\mailsa}\path|{alfred.hofmann, ursula.barth, ingrid.haas, frank.holzwarth,|
\urldef{\mailsb}\path|anna.kramer, leonie.kunz, christine.reiss, nicole.sator,|
\urldef{\mailsc}\path|erika.siebert-cole, peter.strasser, lncs}@springer.com|    
\newcommand{\keywords}[1]{\par\addvspace\baselineskip
\noindent\keywordname\enspace\ignorespaces#1}

\def\ins-figure#1#2#3{
\begin{figure}[t]
 \begin{center}
  \includegraphics[width=#3,keepaspectratio,clip]{./#1.eps}
  \caption{#2}
  \label{fig:#1}
 \end{center}
\end{figure}
}

\def\fig#1{Fig.~\ref{fig:#1}}

\def\Fig#1{Figure~\ref{fig:#1}}
\def\eq#1{Eq.~(\ref{eq:#1})}
\def\eqs#1#2{Eqs.~(\ref{eq:#1}) and (\ref{eq:#2})}

\begin{document}

\mainmatter  

\title{%
A Time-domain Analog\\
Weighted-sum Calculation Model \\
for Extremely Low Power VLSI Implementation\\
of Multi-layer Neural Networks
}

\titlerunning{A Time-domain Weighted-sum Calculation Model}

%
%
\author{Quan Wang%
\and Hakaru Tamukoh \and Takashi Morie}
\authorrunning{Quan Wang \and Hakaru Tamukoh \and Takashi Morie}

\institute{Graduate School of Life Science and Systems Engineering,\\
Kyushu Institute of Technology\\
2-4, Hibikino, Wakamatsu-ku, Kitakyushu, 808-0196 Japan\\
}

\maketitle

\begin{abstract}
A time-domain analog weighted-sum calculation model is proposed based on an
integrate-and-fire-type spiking neuron model.
The proposed calculation model is applied to multi-layer feedforward networks, in which 
weighted summations with positive and negative weights are separately performed in each layer
and summation results are then fed into the next layers without their subtraction operation.
We also propose very large-scale integrated (VLSI) circuits to implement the proposed model.
Unlike the conventional analog voltage or current mode circuits, 
the time-domain analog circuits use transient operation in charging/discharging processes to capacitors.
Since the circuits can be designed without operational amplifiers, 
they can operate with extremely low power consumption.
However, they have to use very high resistance devices on the order of G$\rm \Omega$.
We designed a proof-of-concept (PoC) CMOS VLSI chip to verify weighted-sum operation with the same weights
and evaluated it by post-layout circuit simulation using 250-nm fabrication technology.
High resistance operation was realized by using the subthreshold operation region of MOS transistors.
Simulation results showed that energy efficiency for the weighted-sum calculation was 290~TOPS/W,
more than one order of magnitude higher than that in state-of-the-art 
digital AI processors, even though the minimum width of interconnection used in the PoC chip 
was several times larger than that in such digital processors. 
If state-of-the-art VLSI technology is used to implement the proposed model, 
an energy efficiency of more than 1,000~TOPS/W will be possible. For practical applications, 
development of emerging analog memory devices such as ferroelectric-gate FETs is necessary.


\end{abstract}


\section{Introduction}

Artificial neural networks (ANNs), such as convolutional deep
neural networks (CNNs)~\cite{lecun98} and 
fully-connected multi-layer perceptrons (MLPs)~\cite{ciresan2010}, 
have shown excellent performance on various tasks 
including image recognition~\cite{ciresan2010,alex2012,farabet2013,inception2015,lecun2015deep}.
However, computation in ANNs is very heavy, 
which leads to high power consumption in current digital computers, 
and even in highly parallel coprocessors such as graphics processing units (GPUs). 
In order to implement ANNs at edge devices such as mobile phones and personal service robots, 
very low power consumption operation is required.

In ANN models, weighted summation, or multiply-and-accumulate (MAC) operation, 
is an essential and heavy calculation task,
and dedicated complementary metal-oxide-semiconductor (CMOS)
very-large-scale integration (VLSI) processors have been 
developed to accomplish it~\cite{sim16-isscc,moons17-isscc,shin17-isscc,chen18-isscc}.
As an implementation approach other than digital processors, 
use of analog operation in CMOS VLSI circuits is a promising method
for achieving an extremely low power consumption operation of such a calculation 
task~\cite{fick2017analog,lee201624,miyashita16-asscc,mahmoodi2018ultra}.
Although the calculation precision is limited 
due to the non-idealities of analog operation such as noise and device mismatches, 
the neural network models and circuits can be designed to be robust to 
such non-idealities ~\cite{morie94,indiveri2002,guo2017temperature}.
On the other hand, in the research field of ANN models, low-precision neural networks
have been proposed and their comparable performance has been demonstrated, mainly in
applications of image recognition~\cite{binaryconnect2015,hubara2016quantized}.
These models facilitate the development of energy-efficient hardware 
implementations~\cite{miyashita16-asscc}.

The time-domain weighted-sum calculation model was originally proposed based on 
mathematical spiking neuron models inspired by biological neuron behavior~\cite{maass97-nc,maass99-inbook}. 
We have simplified and expanded this calculation model under the assumption of operation in analog circuits
with transient states, and call its VLSI implementation approach 
``Time-domain Analog Computing with Transient states (TACT).''
In contrast to conventional weighted-sum operation in analog voltage or current modes, 
the TACT approach is suitable for much lower power consumption operation in CMOS VLSI implementation of ANNs.

We have already proposed a device and circuit that performs 
time-domain weighted-sum calculation~\cite{morie10-iscas,tohara16-apex,morie2016spike}. 
The proposed circuit consists of plural input resistive elements 
and a capacitor (RC circuit), which can achieve extremely low-power operation.
The energy consumption could be lowered to the order of 1~fJ per operation, 
which is almost comparable to the calculation efficiency in the brain. 
We also proposed a circuit architecture to implement a weighted-sum calculation with different-signed weights
with two set of RC circuits, one of which calculates positively weighted-sums 
while the other calculates negatively weighted-sums~\cite{wang16-iconip}. 

In this paper, we formulate the weighted-sum calculation model based on the TACT approach,
and propose its applications to ANNs such as MLPs and CNNs.
We also show simulation results of ANNs using the MNIST database 
in which the calculation results by the proposed model are compared with 
the ordinary numerical calculation results, and verify the usefulness of our model.
We then evaluate the energy consumption of the proposed circuit
by conducting post-layout circuit simulation of a CMOS circuit designed equivalently to the RC circuit. 
We propose a VLSI circuit architecture for ANNs based on the proposed model.

\section{Spike-based Time-domain Weighted-sum Calculation Model}

\subsection{Time-domain Weighted-sum Calculation with Same-Signed Weights}

A simple spiking neuron model, also known as an integrate-and-fire-type (IF) neuron
model, is shown in \fig{IFneuron-psp}~\cite{maass99-book}.
In this model, a neuron receives spike pulses via synapses. A
spike pulse only indicates the
input timing, and its pulse width and amplitude 
do not affect the following processing.
A spike generates a temporal voltage change, which is called a post-synaptic
potential (PSP), and the internal potential of the $n$-th neuron,
$V_n(t)$, is equal to the spatiotemporal summation of all PSPs.
When $V_n(t)$ reaches the firing threshold $\theta$, the neuron outputs a spike,
and $V_n(t)$ then settles back to the steady state. 

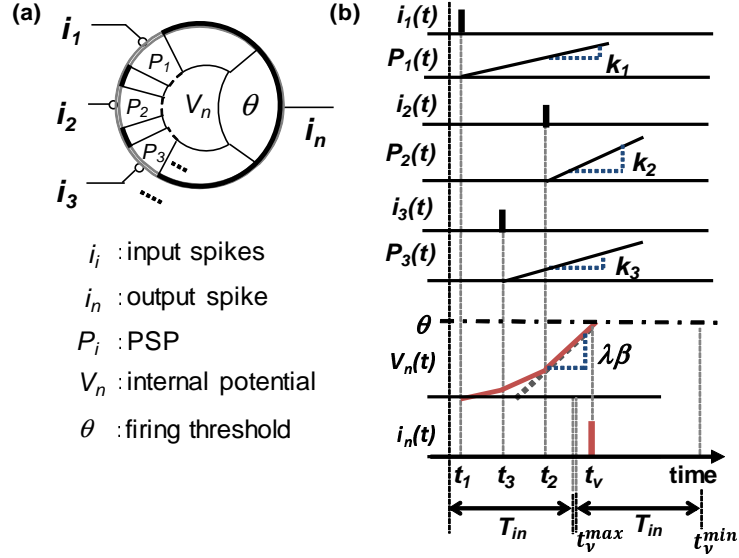
\ins-figure{IFneuron-psp}{IF neuron model for weighted-sum operation:
schematic of the model and weighted-sum operation using rise timing of PSPs.
}{100mm}

Based on the model proposed in~\cite{maass97-nc}, 
a simplified weighted-sum operation model using IF neurons is proposed.
Time span $T_{in}$ is defined, during which only one spike is fed from each neuron, 
and it is assumed that a PSP generated by a spike from neuron $i$
increases linearly with slope $k_{i}$ from the timing of
the spike input, $t_i$, as shown in \fig{IFneuron-psp}. 

A required weighted-sum operation is that normalized variables
$x_i \ (0 \le x_i \le 1, i=1,2,\cdots,N)$ is multiplied
by weight coefficients $a_i$,
and the multiplication results are summed regarding $i$, 
where $N$ is the number of inputs.
This weighted-sum operation can be performed using 
the rise timing of PSPs in the IF neuron model.
Input spike timing $t_i$ is determined based on $x_i$ using the following relation;
\begin{eqnarray}
t_i &=& T_{in}(1-x_i),\\
x_i &=& (1-\frac{t_i}{T_{in}}).
\label{eq:ti}
\end{eqnarray}
Coefficients $a_i$ are transformed into the PSPs' slopes $k_i$;
\begin{equation}
k_i= \lambda a_i, 
\label{eq:ki}
\end{equation}
where $\lambda$ is a positive constant. 
If the firing time of the neuron is defined as
$t_{\nu}$, the following equation is easily obtained:
\begin{equation}
\sum_{i=1}^{N} k_i (t_{\nu} - t_i) = \theta.
\label{eq:mac0}
\end{equation}
If we define the following parameters:
\begin{equation}
\beta = \sum_{i=1}^{N} a_i,
\end{equation}
we obtain
\begin{eqnarray}
\sum_{i=1}^{N} a_i \cdot x_i &=& \frac{\theta/\lambda+ \beta (T_{in}-t_{\nu})}{T_{in}},\label{eq:mac-result}\\
&=& \frac{\theta}{\lambda T_{in}}+ \beta (1-\frac{t_{\nu}}{T_{in}}).\label{eq:mac-result1}
\end{eqnarray}

Here, we assume that all the weights in the calculation have the same sign; 
i.e., $a_i \ge 0$ or $a_i \le 0$ for all $i$. 
When all inputs are minimum ($\forall i\  x_i=0$), 
the left side of \eq{mac-result} is zero. Then, the output timing $t_{\nu}$ is given by
\begin{equation}
t_{\nu}^{min} = \frac{\theta}{\lambda\beta}+T_{in}.
\label{eq:tv0}
\end{equation}
On the other hand, when all inputs are maximum ($\forall i\  x_i=1$), 
the left side of \eq{mac-result} is $\beta$, and the output timing $t_{\nu}$ is given by
\begin{equation}
t_{\nu}^{max} = \frac{\theta}{\lambda\beta}.
\label{eq:tv1}
\end{equation}
The time span during which $t_{\nu}$ can be output is 
$[t_{\nu}^{max}, t_{\nu}^{min}]$, and its interval is
\begin{equation}
T_{out} \equiv t_{\nu}^{min}-t_{\nu}^{max} = T_{in}.
\label{eq:tv01}
\end{equation}
Thus, the time span of output spikes is the same as that of
input spikes, $T_{in}$. 

In this model, since the normalization of the sum of $a_i$ ($\beta=1$) is not required 
(unlike in our previous work~\cite{maass97-nc,maass99-inbook,tohara16-apex}),
the calculation process becomes much simpler.
When implementing the time-domain weighted-sum operation, setting
the threshold potential $\theta$ properly is the key to making the operation
work appropriately. As shown in \fig{IFneuron-psp}, the earliest output spike timing
has to be later than the latest input spiking timing $T_{in}$;
that is, $t_{\nu}^{max} \ge T_{in}$.
Thus, 
 \begin{equation}
\theta \ge \lambda\beta T_{in}.
\label{eq:th01}
 \end{equation}
Also, we can rewrite \eq{th01} as 
\begin{eqnarray}
\theta = \lambda\beta T_{in} + \delta,
\label{eq:thde}\\
\delta = \epsilon(\lambda\beta T_{in}),
\label{eq:deep}
 \end{eqnarray}
where $\epsilon \ge 0$ is an arbitrarily small value. 
By substituting \eqs{thde}{deep} into \eqs{tv0}{tv1}, we obtain
\begin{eqnarray}
t_{\nu}^{min} = (2+\epsilon)T_{in}
\label{eq:tvep0},\\
t_{\nu}^{max} =  (1+\epsilon)T_{in}
\label{eq:tvep1},
 \end{eqnarray}
where $\epsilon T_{in}$ is considered as a time slot between input and output timing spans, 
as shown in \fig{IFneuron-psp}, 
and $\epsilon$ determines the length of the slot.
Also, the weighted summation expressed by \eq{mac-result} is rewritten as follows:
\begin{equation}
\sum_{i=1}^{N} a_i \cdot x_i = \beta [2 + \epsilon -\frac{t_{\nu}}{T_{in}}].\label{eq:mac-result3}
\end{equation}

\subsection{Time-domain Weighted-sum Calculation with Different-Signed Weights}

We have proposed a time-domain weighted-sum calculation model with two spiking neurons,
one of which is for all the positive weights and the
other for all the negative ones~\cite{wang16-iconip}. 
We apply \eq{mac-result} to each neuron,
and the two results are summed as the final result of the original weighted-sum. 
Here, we show the details of the model.

Let $a_{i}^{+}$ and $a_{i}^{-}$ indicate the positive and negative weights, respectively. 
We define
\begin{eqnarray}
\beta_{}^{+}=\sum_{i=1}^{N_{}^{+}}a_{i}^{+} \ge 0,\\
\beta_{}^{-}=\sum_{i=1}^{N_{}^{-}}a_{i}^{-} \le 0.
\end{eqnarray}
where $N_{}^{+}$ and $N_{}^{-}$ are the numbers of positive and negative weights, respectively; 
\begin{eqnarray}
N &=& N_{}^{+}+N_{}^{-},\\
\sum_{i=1}^{N}a_{i} &=& \sum_{i=1}^{N_{}^{+}}a_{i}^{+}+\sum_{i=1}^{N_{}^{-}}a_{i}^{-},\\
\beta &=& \beta_{}^{+}+\beta_{}^{-}.
\end{eqnarray}
Thus, assuming $\lambda=1$, \eq{mac0} is rewritten for the positive and negative weighted-sum operations as follows:
\begin{eqnarray}
\sum_{i=1}^{N_{}^{+}} a_i^{+} (t_{\nu}^{+} - t_i) = \theta_{}^{+}
\label{eq:mac0-pos},\\
\sum_{i=1}^{N_{}^{-}} a_i^{-} (t_{\nu}^{-} - t_i) = \theta_{}^{-}
\label{eq:mac0-nega},
\end{eqnarray}
where $\theta_{}^{+}(>0), \theta_{}^{-}(<0), and t_{\nu}^{+}$ and $t_{\nu}^{-}$ indicate the threshold values 
and output timing for the positively and negatively weighted-sum operation, respectively. Then we obtain
\begin{eqnarray}
\sum_{i=1}^{N_{}^{+}} a_{i}^{+} \cdot x_i &=& \frac{\theta_{}^{+}+ \beta_{}^{+} (T_{in}-t_{\nu}^{+})}{T_{in}}
\label{eq:mac-result-pos},\\
\sum_{i=1}^{N_{}^{-}} a_{i}^{-} \cdot x_i &=& \frac{\theta_{}^{-}+ \beta_{}^{-} (T_{in}-t_{\nu}^{-})}{T_{in}}.
\label{eq:mac-result-nega}
\end{eqnarray}
Therefore, we can obtain the original weighted-sum result:
\begin{eqnarray}
\sum_{i=1}^{N} a_i\cdot x_i &=& \sum_{i=1}^{N_{}^{+}} a_{i}^{+} \cdot x_i + \sum_{i=1}^{N_{}^{-}} a_{i}^{-} \cdot x_i,\\
&=& \frac{\theta_{}^{+}+\theta_{}^{-}+\beta T_{in} -(\beta_{}^{+}t_{\nu}^{+}+\beta_{}^{-}t_{\nu}^{-})}{T_{in}}
\label{eq:pos-nega-in}.
\end{eqnarray}

Let us define a dummy weight $a_0$ as the difference between both absolute values of $\beta_{}^{\pm}$;
\begin{equation}
a_0=-(\beta_{}^{+}+\beta_{}^{-}).
\end{equation}
If $\beta_{}^{+} \ge -\beta_{}^{-}$, then $a_0 \le 0$ and this dummy weight is incorporated into 
the negative weight group, and vice versa. This dummy weight is related to a zero input,
$x_0=0$, which means $t_0=T_{in}$.
By using the dummy weight, we can make the absolute values of $\beta_{}^{\pm}$ identical
($\beta=0$), and we define
\begin{equation}
\beta_o = \beta_{}^{+}=-\beta_{}^{-}
\label{eq:unif-beta}.
\end{equation}
Also, according to \eqs{thde}{deep}, the absolute values of $\theta_{}^{+}$ and $\theta_{}^{-}$ 
can be the same, and $\theta_{}^{+}+\theta_{}^{-}=0$. Therefore, \eq{pos-nega-in} can be rewritten as
\begin{equation}
\sum_{i=1}^{N} a_i\cdot x_i=\frac{\beta_o(t_{\nu}^{-}-t_{\nu}^{+})}{T_{in}}
\label{eq:final-ws}.
\end{equation}

\section{Spiking Neural Network model}

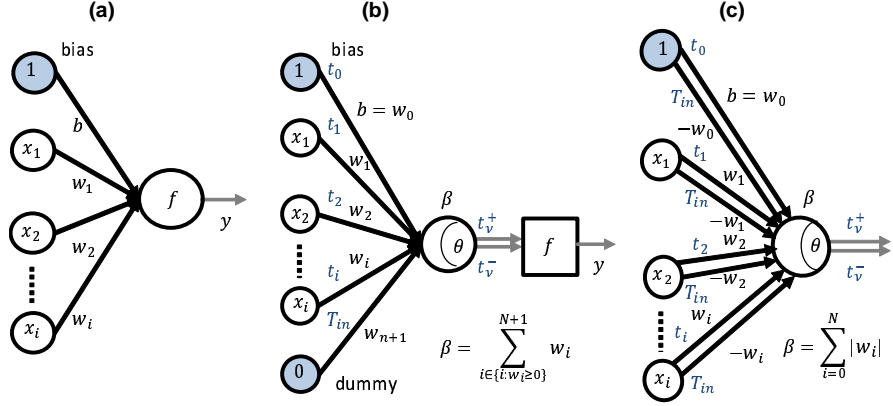
\ins-figure{NeuronModel}{Neuron model: (a) typical neuron model; (b) neuron model for time-domain weighted-sum 
operation with a dummy weight, $w_{n+1}$; (c) neuron model for time-domain weighted-sum operation in which each
synapse has two sets of inputs and weights. One is $(x_i, w_i)$ and the other is $(0, -w_i)$ 
or $(t_i , w_i)$ and $(T_{in}, -w_i)$ according to \eq{ti}.
}{130mm}

\subsection{Neuron Model}

The typical neuron model of artificial neural networks is shown in \fig{NeuronModel}(a),
which has $N$ inputs $x_i$ with weights $w_i$ and a bias $b$;
\begin{equation}
y=f(\sum_{i=1}^{N} w_i \cdot x_i + b),
\end{equation} 
where $y$ is the output of the neuron, and
$f$ is an activation function. 
We can consider the bias as a weight whose input is always unity. 
Therefore, our time-domain weighted-sum calculation model with the dummy weight
can be applied to this neuron model, as shown in \fig{NeuronModel}(b). 
According to \eq{final-ws}, 
\begin{equation}
\sum_{i=1}^{N} w_i \cdot x_i + b=\frac{\beta(t_{\nu}^{-}-t_{\nu}^{+})}{T_{in}}.
\label{eq:tdws-final-rst}
\end{equation}

Based on \eq{tdws-final-rst}, we propose another model, shown in \fig{NeuronModel}(c),
in which each synapse has two sets of inputs and weights; one is $(x_i , w_i)$ and the other is $(0, -w_i)$. 
In this model, it is not necessary to add a dummy weight 
because the summation of positive weights is $\beta=\sum_{i=0}^{N} |w_i|$ 
and that of negative ones is $\beta=-\sum_{i=0}^{N} |w_i|$,
which means that the absolute values of both summations are equal.

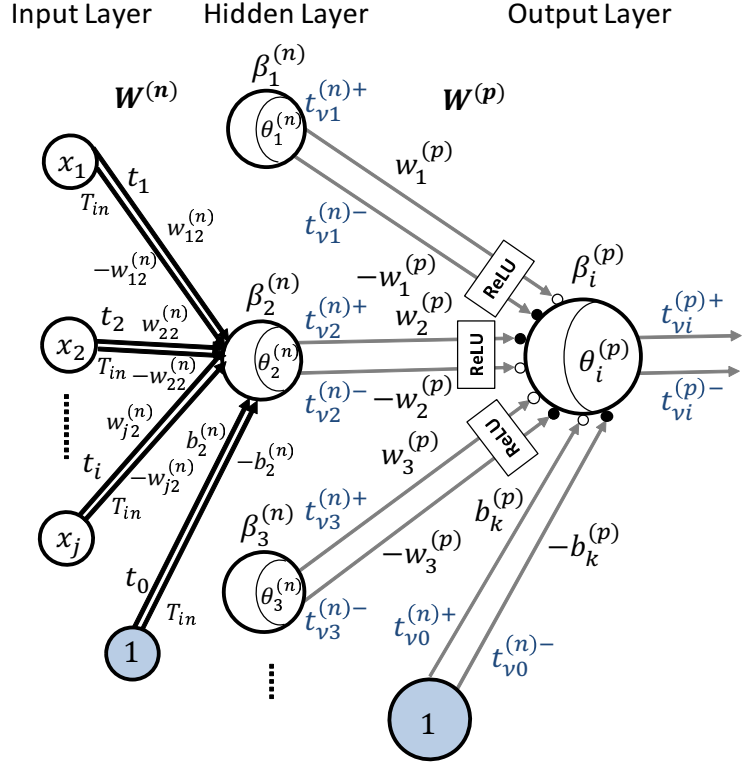
\ins-figure{MLP-Two-IOs}{General neural network model with two inputs and outputs for time-domain
weighted-sum calculation with positive and negative weights.
}{100mm}

As the activation function $f$, we often use the rectified linear unit called ``ReLU''
~\cite{glorot2011,HeK2015}, which is defined as follows:
\begin{eqnarray}
f(x)=ReLU(x)=\cases{
x & if $x \ge 0$,\cr
0 & otherwise.\cr
}
\end{eqnarray}
We can implement the ReLU function by comparing the output timings $t_{vj}^{(n)-}$ and $t_{vj}^{(n)+}$ 
in the time-domain weighted-sum calculation. 
If $t_{vk}^{(n)-} \ge t_{vk}^{(n)+}$, 
the difference between two timing values is regarded as the output 
transferred to neurons in the next layer.
On the other hand, if $t_{vk}^{(n)-} < t_{vk}^{(n)+}$, 
the output is zero, because the total weighted sum is negative.
To do this, we set $t_{vk}^{(n)-}$ and $t_{vk}^{(n)+}$ to be identical.
Its circuit implementation will be shown later.

\subsection{Neural Network Model}

In this section, we show an application of our time-domain weighted-sum model 
to the MLPs shown in \fig{MLP-Two-IOs} as an example that has one hidden layer and 
two sets of input and weight for each neuron.
In this application, after we calculate a weighted sum using \eq{tdws-final-rst}, 
the result is given to the activation function $f$, and the output is fed into the next layer.

According to \eq{tdws-final-rst}, the weighted-sum result of the $j$-th neuron in the layer
labeled $n$ in \fig{MLP-Two-IOs} can be
\begin{equation}
\sum_{i=0}^{N} w_{ij}^{(n)} \cdot x_i = \frac{\beta_j^{(n)}}{T_{in}}(t_{vj}^{(n)-}-t_{vj}^{(n)+}).
\end{equation}
where $w_{0j}^{(n)}=b_j^{(n)}$ is the bias of the $j$-th neuron in the $n$-th layer.
The output of the $k$-th neuron in the layer labeled $p (=n+1)$ in \fig{MLP-Two-IOs} is
\begin{equation}
\sum_{j=1}^{N} w_{jk}^{(p)} \cdot f(\sum_{i=0}^{N} w_{ij}^{(n)} \cdot x_i)+b_k^{(p)} =\sum_{j=1}^{N} w_{jk}^{(p)} \cdot f(\frac{\beta_j^{(n)}}{T_{in}}(t_{vj}^{(n)-}-t_{vj}^{(n)+}))+b_k^{(p)}.
\label{eq:nn-synapse}
\end{equation}
Here, substituting the activation function with ReLU, we can obtain
\begin{eqnarray}
ReLU(\frac{\beta_j^{(n)}}{T_{in}}(t_{vj}^{(n)-}-t_{vj}^{(n)+}))=\frac{\beta_j^{(n)}}{T_{in}}(t_{vj}^{(n)-}-t_{vj}^{(n)+}),
\end{eqnarray}
where if $t_{vk}^{(n)-} < t_{vk}^{(n)+}$, then let $t_{vk}^{(n)-}=t_{vk}^{(n)+}$. 
Thus, \eq{nn-synapse} can be rewritten as
\begin{equation}
\sum_{j=1}^{N} w_{jk}^{(p)} \cdot ReLU(\sum_{i=0}^{N} w_{ij}^{(n)} \cdot x_i)+b_k^{(p)} =\sum_{j=1}^{N} w_{jk}^{(p)} \cdot \frac{\beta_j^{(n)}}{T_{in}}(t_{vj}^{(n)-}-t_{vj}^{(n)+})+b_k^{(p)}.
\label{eq:nn-synapse-relu}
\end{equation}

In the MLP shown in \fig{MLP-Two-IOs}, 
we transfer the output timings $t_{vj}^{(n)+}$ and $t_{vj}^{(n)-}$ generated in layer $n$
to neurons in layer $p$ and perform the time-domain weighted-sum operation.
In layer $n$, we assume that 
timing $t_{vj}^{(n)+}$ is related to weight $w_{jk}^{(p)}$ and 
$t_{vj}^{(n)-}$ is related to $-w_{jk}^{(p)}$. 
We also assume here $j=3$, and that 
$w_{1k}^{(p)} \ge 0, w_{2k}^{(p)} < 0, w_{3k}^{(p)} \ge 0, b_k^{(p)} \ge 0$, 
and $\theta_k^{(p)+}=- \theta_k^{(p)-}$,
where $\theta_k^{(p)+}$ and $\theta_k^{(p)-}$ are the threshold values for positively 
and negatively weighted-sum operations, respectively.
Thus, according to \eq{mac0}, we can obtain
\begin{eqnarray}
w_{1k}^{(p)}(t_{vk}^{(p)+}-t_{v1}^{(n)+})+(-w_{2k}^{(p)})(t_{vk}^{(p)+}-t_{v2}^{(n)-})+w_{3k}^{(p)}(t_{vk}^{(p)+}-t_{v3}^{(n)+})\cr
+b_k^{(p)}(t_{vk}^{(p)+}-t_{v0}^{(n)+})=\theta_k^{(p)+}
\label{eq:mac-bi-pos}\\
(-w_{1k}^{(p)})(t_{vk}^{(p)-}-t_{v1}^{(n)-})+w_{2k}^{(p)}(t_{vk}^{(p)-}-t_{v2}^{(n)+})+(-w_{3k}^{(p)})(t_{vk}^{(p)-}-t_{v3}^{(n)-})\cr
+(-b_k^{(p)})(t_{vk}^{(p)-}-t_{v0}^{(n)-})=\theta_k^{(p)-}
\label{eq:mac-bi-nega}
\end{eqnarray}
By adding \eq{mac-bi-pos} to \eq{mac-bi-nega}, the following relationship is obtained:
\begin{eqnarray}
w_{1k}^{(p)}(t_{vk}^{(p)+}-t_{v1}^{(n)+})+(-w_{2k}^{(p)})(t_{vk}^{(p)+}-t_{v2}^{(n)-})+w_{3k}^{(p)}(t_{vk}^{(p)+}-t_{v3}^{(n)+})+\cr
(-w_{1k}^{(p)})(t_{vk}^{(p)-}-t_{v1}^{(n)-})+w_{2k}^{(p)}(t_{vk}^{(p)-}-t_{v2}^{(n)+})+(-w_{3k}^{(p)})(t_{vk}^{(p)-}-t_{v3}^{(n)-})\cr
+b_k^{(p)}(t_{vk}^{(p)+}-t_{v0}^{(n)+})+(-b_k^{(p)})(t_{vk}^{(p)-}-t_{v0}^{(n)-})\cr
=t_{vk}^{(p)+}(w_{1k}^{(p)}-w_{2k}^{(p)}+w_{3k}^{(p)}+b_k^{(p)})+t_{vk}^{(p)-}(-w_{1k}^{(p)}+w_{2k}^{(p)}-w_{3k}^{(p)}-b_k^{(p)})\cr
   +\sum_{j=0}^{N=3} w_{jk}^{(p)} \cdot t_{vj}^{(n)-}-\sum_{j=0}^{N=3} w_{jk}^{(p)} \cdot t_{vj}^{(n)+}\cr
=\sum_{j=0}^{N=3} |w_{jk}^{(p)}| \cdot (t_{vk}^{(p)+}-t_{vk}^{(p)-}) + \sum_{j=0}^{N=3} w_{jk}^{(p)} \cdot(t_{vj}^{(n)-}-t_{vj}^{(n)+})=0,
\end{eqnarray}
where $w_{0k}^{(p)}=b_k{(p)}$.
Thus, we can obtain the following simple expression:
\begin{equation}
\sum_{j=0}^{N=3} w_{jk}^{(p)} \cdot(t_{vj}^{(n)-}-t_{vj}^{(n)+}) =  (t_{vk}^{(p)-}-t_{vk}^{(p)+})\cdot \sum_{j=0}^{N=3} |w_{jk}^{(p)}|.
\end{equation}
Therefore, we can have the more general expression as follows:
\begin{equation}
\sum_{j=0}^{N} w_{jk}^{(p)} \cdot(t_{vj}^{(n)-}-t_{vj}^{(n)+}) = (t_{vk}^{(p)-}-t_{vk}^{(p)+})\cdot \sum_{j=0}^{N} |w_{jk}^{(p)}| .
\end{equation}
Then, we can obtain the following formula:
\begin{eqnarray}
\sum_{j=0}^{N} w_{jk}^{(p)} \cdot \frac{\beta_j^{(n)}}{T_{in}}\cdot(t_{vj}^{(n)-}-t_{vj}^{(n)+}) = 
(t_{vk}^{(p)-}-t_{vk}^{(p)+})\cdot \sum_{j=0}^{N} |w_{jk}^{(p)}|\cdot \frac{\beta_j^{(n)}}{T_{in}},\\
\sum_{j=1}^{N} w_{jk}^{(p)} \cdot \frac{\beta_j^{(n)}}{T_{in}}\cdot(t_{vj}^{(n)-}-t_{vj}^{(n)+})+w_{0k}^{(p)}\cdot 
\frac{\beta_0^{(n)}}{T_{in}}(t_{v0}^{(n)-}-t_{v0}^{(n)+}) \cr
= (t_{vk}^{(p)-}-t_{vk}^{(p)+})\cdot \sum_{j=0}^{N} |w_{jk}^{(p)}|\cdot \frac{\beta_j^{(n)}}{T_{in}},
\label{eq:mac-continuous}
\end{eqnarray}
where $w_{0k}^{(p)}=b_k^{(p)}$, $t_{v0}^{(n)-}=T_{in}$, and $t_{v0}^{(n)+}=0$. 
Because there is no input to the bias $b_k^{(p)}$,
we let $\beta_0^{(n)}=1$. Therefore, \eq{mac-continuous} becomes
\begin{equation}
\sum_{j=1}^{N} w_{jk}^{(p)} \cdot \frac{\beta_j^{(n)}}{T_{in}}\cdot(t_{vj}^{(n)-}-t_{vj}^{(n)+})+b_k^{(p)} = 
(t_{vk}^{(p)-}-t_{vk}^{(p)+})\cdot \sum_{j=0}^{N} |w_{jk}^{(p)}|\cdot \frac{\beta_j^{(n)}}{T_{in}},
\end{equation}
where $\beta_0^{(n)}=1$ on the right side.
Therefore, \eq{nn-synapse-relu} can be modified as
\begin{eqnarray}
\sum_{j=1}^{N} w_{jk}^{(p)} \cdot ReLU(\sum_{i=0}^{N} w_{ij}^{(n)} \cdot x_i)+b_k^{(p)}=
(t_{vk}^{(p)-}-t_{vk}^{(p)+})\cdot \sum_{j=0}^{N} |w_{jk}^{(p)}|\cdot \frac{\beta_j^{(n)}}{T_{in}} .
\label{eq:nn-synapse-timing}
\end{eqnarray}

As a result, for neurons in the hidden layer $n$, we apply the time-domain weighted-sum operation 
to generate the timing $t_{vj}^{(n)+}$ and $t_{vj}^{(n)-}$ 
for the positively and negatively weighted-sum calculation from the input layer, respectively. 
Then, these timings are directly transferred to neurons in the next layer $p$,
and timing $t_{vj}^{(p)+}$ and $t_{vj}^{(p)-}$ are obtained.
Finally, we calculate the final outputs of the MLP using \eq{nn-synapse-timing}. 
Note that intermediate weighted-sum results with different-signed weights are not calculated in the middle layers. 

For CNNs, weighted-sum calculations of convolutions can be performed in the same way.
In addition to the convolutions, max pooling operations can also be implemented simply by considering the 
difference between positive and negative timing values, as follows:

\begin{equation}
y_{maxpooling}=(t_{vk}^{(l)-}, t_{vk}^{(l)+}),
\label{eq:max-pooling}
\end{equation}
where 
\begin{equation}
k = \arg \max_i (t_{v1}^{(l)-}-t_{v1}^{(l)+},\cdots, t_{vi}^{(l)-}-t_{vi}^{(l)+}).
\label{eq:max-pooling1}
\end{equation}

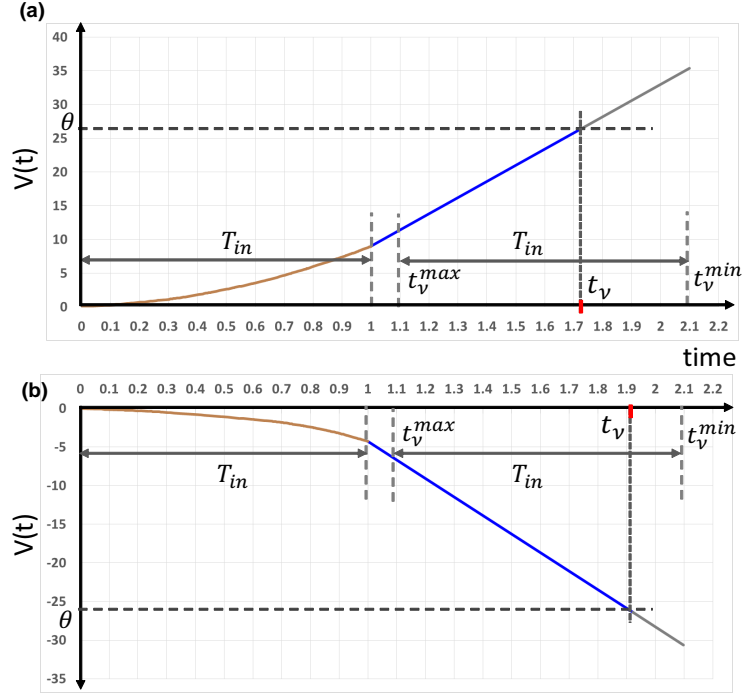
\ins-figure{Result-psp}{Simulation results for the time-domain weighted-sum calculation model:
(a) PSP of positively weighted-sum operation with 249 inputs in which 
$T_{in}=1,\lambda=1,\beta^{+}=24.01$, and $\theta^{+}=26.41$.
The output spike timing is $t_{\nu}^{+}=1.7256$.
(b) PSP of negatively weighted-sum operation with 253 inputs in which 
$w_0=-0.06,w_{n+1}=-2.819,T_{in}=1, \lambda=1,\beta^{-}=-24.01$, and $\theta^{-}=-26.41$.
The output spike timing is $t_{\nu}^{-}=1.9221$.
Thus, the result of weighted-sum calculation
is $|\beta^{\pm}|(t_{\nu}^{-}-t_{\nu}^{+})/T_{in}=4.718$. 
}{100mm}

\subsection{Numerical Simulations of Neural Networks}

We performed numerical simulations to verify our weighted-sum calculation model. 
First, in order to verify our model for weighted-sum calculation with different-signed weights,
we conducted a simulation to perform a weighted-sum calculation 
with 501 pairs of inputs and weights that consisted of 249 positive and 252 negative weights.
We added a dummy weight to make the sum of positive weights equal to the absolute sum of the negative ones.
\Fig{Result-psp} shows the simulation results of time-domain weighted-sum calculation with a dummy weight $w_{n+1}$. 
The results show that the weighted-summation can be calculated correctly with different
negative and positive firing timing inputs each set of which are multiplied by the corresponding signed weights.

Then, we applied our model to a four-layer MLP (784-100-100-10) and a CNN known as LeNet5~\cite{lecun98} 
to classify the MNIST digit character set. We trained these two ANNs, and then performed inference according to \eq{nn-synapse-timing}
with the obtained weights, which were either binary~\cite{binaryconnect2015} or analog values.
As described above, output spike timing at each neuron in the previous layer was directly conveyed to the neurons in the next layer 
without obtaining the subtraction of the signed weighted-sum results.
We founded that we obtained the same weighted-sum calculation results in the last layer, and also the same
recognition precisions in both NNs as in the numerically calculated ones.

\section{Circuits and Architectures for TACT-based Neural Networks}

As an implementation for our weighted-sum calculation based on our TACT approach,
we have proposed an RC circuit in which a capacitor is connected by plural resistors, as shown in \fig{synapse-circuit}(a).
Theoretical estimations have indicated that this circuit can perform weighted-sum calculations 
with extremely low energy consumption~\cite{tohara16-apex,wang16-iconip}.
In CMOS VLSI implementation, resistance $R$ can be replaced by a p-type MOS field-effect transistor (pMOSFET), 
as shown in \fig{synapse-circuit}(b).
The approximately linear slope $k$ is generated by capacitance $C$ and ON resistance of a pMOSFET with a step voltage input
$V_{in}$, where we use step voltages instead of spike pulses as inputs.
Each resistance should have a rectification function to prevent an inverse current.

The rectification function is automatically realized by the FET operation as follows. 
When a pMOSFET receives a step-voltage input, the terminal voltage of the input is higher than that at $C$, 
and therefore the input-side terminal of the pMOSFET is ``source'', and the capacitor-side terminal is ``drain.''
In this state, if the gate-source voltage of the pMOSFET is set to exceed its threshold voltage, 
the pMOSFET turns on, and $C$ is charged up.
On the other hand, when a pMOSFET receives no input, the terminal voltage of the input is lower than that at $C$, 
and therefore the source-drain position in the pMOSFET is reversed; 
i.e., the input-side terminal of the pMOSFET is ``drain'', and the capacitor-side terminal is ``source.''
In this state, if the gate-source voltage of the pMOSFET is set not to exceed its threshold voltage, 
the pMOSFET turns off, and charges stored at $C$ does not flow back to the input side.

In order to evaluate the energy consumption of this circuit, we designed a CMOS circuit equivalent to the RC circuit. 
It is obviously difficult to change the ON resistance of each pMOSFET independently, 
because different analog voltages have to be given as the gate voltages of the pMOSFETs.
Therefore, in the PoC circuit, all MOSFETs have the same ON resistance with the same gate voltage.
To realize different analog weights, it is necessary to use 
analog memory devices such as resistance-change memory~\cite{prezioso15-nature,kim15-iedm},
ferroelectric-gate FETs~\cite{li2009threshold},
and floating-gate flash memory~\cite{bavandpour2017energy}.

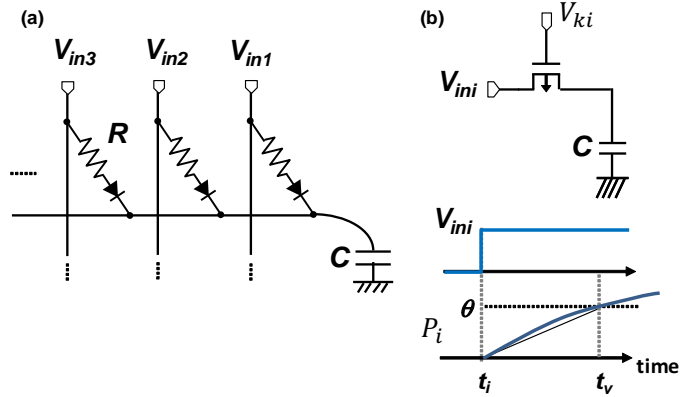
\ins-figure{synapse-circuit}{Synapse circuit: (a) step voltage input and a resistance-capacitance (R-C) circuit 
in which a pMOSFET acts as resistance $R$, and parasitic capacitance of interconnection and 
the gate capacitance of MOSFETs act as $C$ in a VLSI circuit;
(b) approximately linear response of the step voltage input at timing $t_i$ with a slope determined by gate voltage $V_{ki}$.
}{90mm}

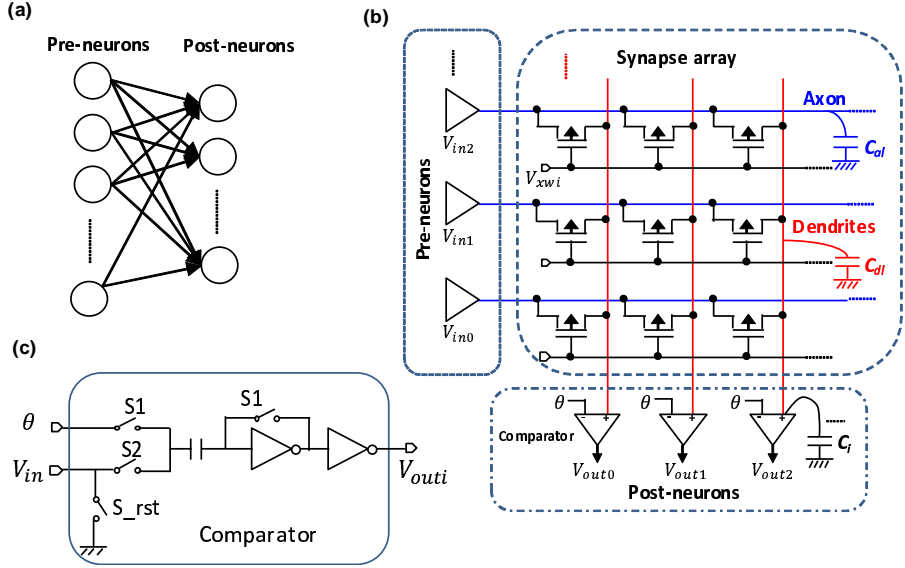
\ins-figure{CrossbarNN}{Crossbar architecture: (a) one-layer neural network model;
(b) crossbar architecture with pre-neurons having axons, post-neurons having dendrites;
(c) CMOS comparator that act as a post-neuron circuit.
}{120mm}

We designed a crossbar synapse circuit array to perform the weighted-sum calculation shown in \fig{IFneuron-psp} 
with a one-layer MLP model, as shown in \fig{CrossbarNN}. 
In the array, the horizontal and vertical lines are referred to as 
``axons'' of the previous neurons and ``dendrites'' of the post neurons, respectively;
each axon line has $M$ synapse circuits, and each dendrite line receives $N$ synapse outputs.
An input voltage charges up the parasitic capacitance of the axon line, $C_{al}$,
and then charges up the capacitance, which includes the parasitic capacitance of the dendrite lines, $C_{dl}$,
and the input capacitance of the post neuron, $C_{i}$, via synapse blocks, each of which consists of a pMOSFET.

We designed two single-layer neural network circuits, which have 500 inputs and 20 outputs ($N = 500, M = 20$), 
and 500 inputs and 100 outputs ($N = 500, M = 100$), respectively.
The layout result and post layout simulation results are shown in \fig{PostLayout}.
The fabrication technology of TSMC 250-~nm were used, and 
both the gate length and width of pMOSFETs were 0.6~$\mu$m.
The simulation results show a correct weighted-sum operation, where $T_{in} = 1~\mu$s, and $\theta = 0.3$~V.

We extracted the parasitic capacitance of the dendrite and axon lines per synapse, 
$c_{dl} = C_{dl}/N$ and $c_{al} = C_{al}/M$, which were 1.76~fF and 1.78~fF, respectively. 
It is assumed that the input capacitance of pMOSFETs is included in this capacitance.
Therefore, the energy consumption of line charge/discharge operation 
per synapse, $E_{syn}$, is expressed by
$E_{syn}=(c_{dl}+c_{al})V_{dd}^2$, and it was 3.54~fJ, where $V_{dd}=1$~V. 
We note that the above estimation does not include the energy consumption related to 
charging/discharging of the input capacitance $C_{i}$ of post-neuron circuits;
this, however, would be negligible 
compared to the dendrite line capacitance under 250-nm technology.

As for the neuron part, which consists of a comparator and the output buffer, the energy
consumption was about 1.67~pJ, which means 3.34~fJ per synapse operation. 
As a result, overall energy consumption was 6.88~fJ per synapse operation, 
which consists of two operations, multiply and accumulate.
This implies that the energy efficiency is 290~TOPS/W (Tera-Operations Per Second per Watt). 
This efficiency is more than one order of magnitude higher than that of state-of-the-art 
digital AI processors~\cite{chen18-isscc}.

For sufficient calculation precision, we expect the time constant of RC circuits to be much more than 
1~$\mu$s to guarantee a time resolution of 7 bits, assuming a resolved time step of 10~ns.
To obtain this time constant, $R$ should be more than 1~G$\rm \Omega$, which means that 
the current flowing through each resistance is less than 1~nA.
Such high resistance can be achieved by using the subthreshold operation of MOSFET, and
we set all the gate-source voltages of the pMOSFETs at -0.37V.

We propose the circuit architecture of a neural network suitable for our TACT approach,
in which the weights may be both positive and negative as described above.
The architecture is shown in \fig{TwoLayerNN-archi}, and is composed of synapses acting as resistive elements,
the neuron part functioning as thresholding, the ReLU part denoting the activation, and the configuration part
controlling synapses.

There are two inputs for each synapse circuit, which are $t_i$ as signal input and $T_{in}$
as a dummy input in the first layer, and $t_{\nu i}^{+}$ and $t_{\nu i}^{-}$ at the subsequent layers.
Pairs of positive and negative timing are directly connected to the next layer without calculating subtraction
between positively and negatively signed weighted results.
The synapse part is designed with two resistive elements and two pairs of switches. 
A set of two identical resistances represents the weight value.
We can assume that the upper-side axon is for $t_{\nu i}^{+}$ and the other is for $t_{\nu i}^{-}$, 
and the left-side dendrite is for a positive weight connection while the other is for a negative one.
The two switches are exclusively controlled according the corresponding sign of weights,
which is controlled by the weight control circuit.

The neuron part includes a comparator. The ReLU activation function can easily be implemented by logic gates,
as shown in \fig{ReLU-circuit}. When  $t_{\nu i}^{+} > t_{\nu i}^{-}$, we set both timing at 
$t_{\nu i}^{+}$ as shown in \fig{ReLU-circuit}. 
With such circuits, the output spike timings at each neuron in the previous layer are 
directly transferred into the neurons in the next layer, 
while the nonlinear activation function ReLU can be implemented with low energy consumption operation.

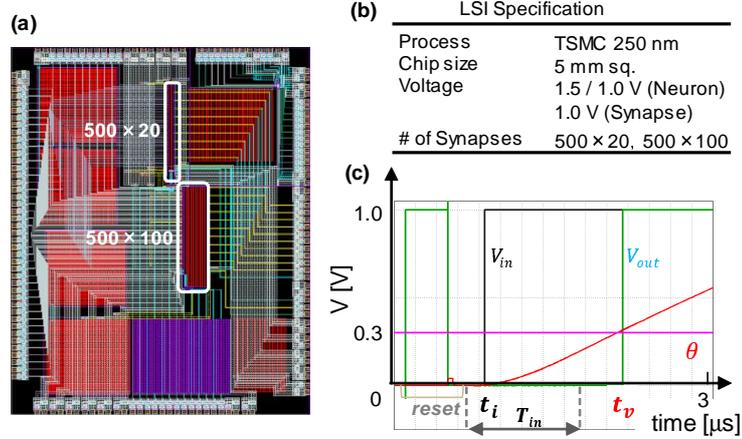
\ins-figure{PostLayout}{Layout results: (a) layout of 500-20 and 500-100 MLPs;
(b) VLSI circuit specification; and (c) post-layout simulation results for time-domain analog calculation.
}{100mm}

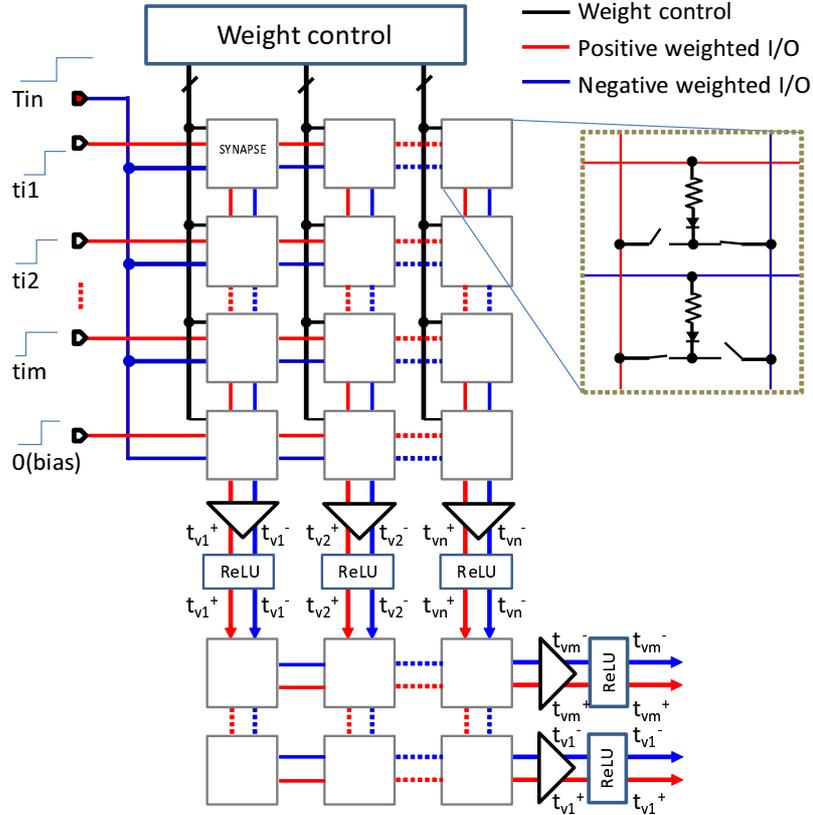
\ins-figure{TwoLayerNN-archi}{Two-layer MLP architecture.
}{110mm}

\ins-figure{ReLU-circuit}{ReLU circuit.
}{80mm}

\section{Conclusions}

In this paper, we proposed a time-domain weighted-sum calculation model based
on the spiking neuron model, and formulated the calculation model 
to implement MLPs with an activation function of ReLU.
In the proposed model, weighted-sum results with different-signed weights 
are not calculated in intermediate layers.
We also proposed VLSI circuits based on the TACT approach
to implement the calculation model with extremely low energy consumption.
We demonstrated the high energy efficiency of the circuit using 250-nm CMOS VLSI technology. 
If we use a more advanced VLSI fabrication technology, able to achieve lower parasitic capacitance,
the energy efficiency will be much improved over 1~POPS/W (Peta-operations Per Second per Watt).

However, there are some issues to be overcome toward developing practical AI processors using our TACT approach.
In terms of synapse circuits, the designed PoC CMOS VLSI chip provided no memory function. 
In order to construct neural network systems, it is necessary to introduce analog memory devices with high resistance
in synapse circuits.
As for the neuron parts, the results of post-layout simulations suggest that 
the energy consumption of this part is comparable to that of the whole synapse part with 500 inputs. 
It will be necessary to design a comparator with much lower power consumption to improve the energy efficiency of 
the whole weighted-sum calculation circuit.

\subsubsection*{Acknowledgments.}

This work was supported by JSPS KAKENHI Grant Nos. 22240022 and 15H01706. 
Part of the work was carried out under a project commissioned by the New Energy and Industrial Technology Development Organization (NEDO),
and the Collaborative Research Project of the Institute of Fluid Science, Tohoku University.



\end{document}